# Magnonic band structure in CoFeB/Ta/NiFe meander-shaped magnetic bilayers

G. Gubbiotti,[1] A. Sadovnikov,[2] E. Beginin,[2] S. Sheshukova,[2] S. Nikitov,[2,3,4] G. Talmelli,[5,6] I. Asselberghs,[5] I. P. Radu,[5] C. Adelmann,[5] and F. Ciubotaru[5]

[1] Istituto Officina dei Materiali del CNR (CNR-IOM), Sede Secondaria di Perugia, c/o Dipartimento di Fisica e Geologia, Università di Perugia, I-06123 Perugia, Italy.

[2] Laboratory Magnetic Metamaterials, Saratov State University, 83 Astrakhanskaya street, Saratov 410012, Russia

[3] Kotel'nikov Institute of Radioengineering and Electronics, Russian Academy of Sciences, Moscow 125009, Russia

[4] Terahertz Spintronics Laboratory, Moscow Institute of Physics and Technology (National University), 9 Institutskii lane, Dolgoprudny, 141700, Russia

[5] Imec, 3001 Leuven, Belgium

[6] KU Leuven, Departement Materiaalkunde, 3001 Leuven, Belgium

**ABSTRACT**

In this work, we investigate the spin-wave propagation in three-dimensional nanoscale CoFeB/Ta/NiFe meander structures fabricated on a structured $SiO_2$/Si substrate. The magnonic band structure has been experimentally determined by wavevector-resolved Brillouin light scattering (BLS) spectroscopy and a set of stationary modes interposed by two dispersive modes of Bloch type have been identified. The results could be understood by micromagnetic and finite element simulations of the mode distributions in both real space and the frequency domain. The dispersive modes periodically oscillate in frequency over the Brillouin zones and correspond to modes, whose spatial distributions extend over the entire sample and are either localized exclusively in the CoFeB layer or the entire CoFeB/Ta/NiFe magnetic bilayer. Stationary modes are mainly concentrated in the vertical segments of the CoFeB and NiFe layers and show negligible amplitudes in the horizontal segments. The findings are compared with those of single-layer CoFeB meander structures with the same geometry parameters, which reveals the influence of the dipolar coupling between the two ferromagnetic layers on the magnonic band structure.

**Corresponding author:** * gubbiotti@iom.cnr.it

In the last decade, magnonic crystals (MCs) — magnetic metamaterials with periodically modulated properties at the nanoscale — have attracted much interest because of the possibility to engineer spin-wave (SW) band structures.[1,2,3] Traditional MCs are based on periodic planar 1D and 2D nanostructured materials[4,5,6,7] and have already found application in magnonic devices such as magnetic field sensors,[8] microwave filters,[9] and magnonic transistors.[10] The recent interest in 3D MCs is driven by the potential for vertically integrated magnonic devices that may ultimately mimic the current development of 3D integrated microelectronic circuits.[11,12] Exploiting the third dimension in magnonics might enhance magnon device functionalities with respect to planar structures due to the interplay between interlayer exchange and dipolar interaction.[13] In addition, 3D magnonic systems might offer several advantages over 2D systems, such as permitting more functionality in a smaller space, allowing for a large number of vertical connections between planar layers, or increasing the density of elements for the fabrication of scalable and configurable magnonic networks. Among 3D MCs, meander-shaped ferromagnetic films fabricated on top of pre-patterned substrates have been proposed as prototypes for the transmission of SW signals in 3D magnonic networks.[14,15,16] A very recent experimental investigation of the magnonic band structure in periodically modulated CoFeB films by Brillouin light scattering (BLS) spectroscopy has revealed that the SW dispersion relation exhibits a periodic character with alternating frequency bands, in which SW propagation is either allowed or forbidden.[17]

In this letter, we investigate by Brillouin light scattering spectroscopy (BLS) the dispersion of collective SWs in 3D MCs based on meander structures including $Co_{40}Fe_{40}B_{20}/Ta/Ni_{80}Fe_{20}$ (CoFeB/Ta/NiFe) magnetic bilayers. The unit cell of the meander structures had a height of $h = 50$ nm and periodicity of $a = 600$ nm, leading to a lattice periodicity in reciprocal space (p/$a = 0.52 \times 10^7$ rad/m) and a wavevector range that were accessible by our experimental BLS setup. The spin-wave dispersion relation was mapped up to the fourth Brillouin zone by sweeping the wave vector along the periodicity direction. The measured SW spectra contained a discrete set of dispersionless modes, *i.e.* with frequencies independent of the wavevectors $k$, at low and high frequencies, whereas two dispersive SWs modes were found in the intermediate frequency range. The theoretical SW dispersions were determined by micromagnetic (MuMax3) and finite element (COMSOL) simulations, and the magnonic band and the spatial profiles of the main modes were reconstructed. The results indicated that propagating modes correspond to SWs with a spatial profile extending throughout the structures. Finally, the magnonic band structure and mode profiles were compared to those of identical meander structures comprising a single CoFeB layer.

The samples were processed as follows: in a first step, 300 nm thick thermal $SiO_2$ was grown on a 300 mm Si (100) wafer and patterned into a periodic grating (height $h = 50$ nm, line and trench widths of 300 nm each) by conventional deep-UV photolithography and reactive ion etching. Next, a Ta(2 nm)/ Co$_{40}$Fe$_{40}$B$_{20}$(23 nm)/Ta(2 nm) stack was deposited by physical vapor deposition (PVD) onto the grating, as shown in Fig. 1. The length of the horizontal segments was $L_1 = 300$ nm and $L_2 = 150$ nm, so that the unit cell had a periodicity of $a = L_1 + 2L_2 = 600$ nm, resulting in Brillouin zone (BZ) edge of $p/a = 0.52 \times 10^7$ rad/m. The limited conformality of the PVD process led to approximately half the film thickness on the vertical segments (sidewalls) with respect to the horizontal ones. After air exposure, a Ta layer of 3 nm was deposited onto the structure, leading to a total Ta spacer of $t_1 = 5$ nm ($t_2 \approx 2.5$ nm), followed by the *in-situ* deposition of a 23 nm thick Ni$_{80}$Fe$_{20}$ magnetic layer. An otherwise identical sample without the NiFe layer was also fabricated used as a reference. Figs. 1(b) and (c) show scanning electron microscopy (SEM) images of the CoFeB/Ta/NiFe sample. It can be seen clearly that all layers coat the patterned $SiO_2$ structures and are connected by sidewall segments, despite the limited conformality of the PVD process.

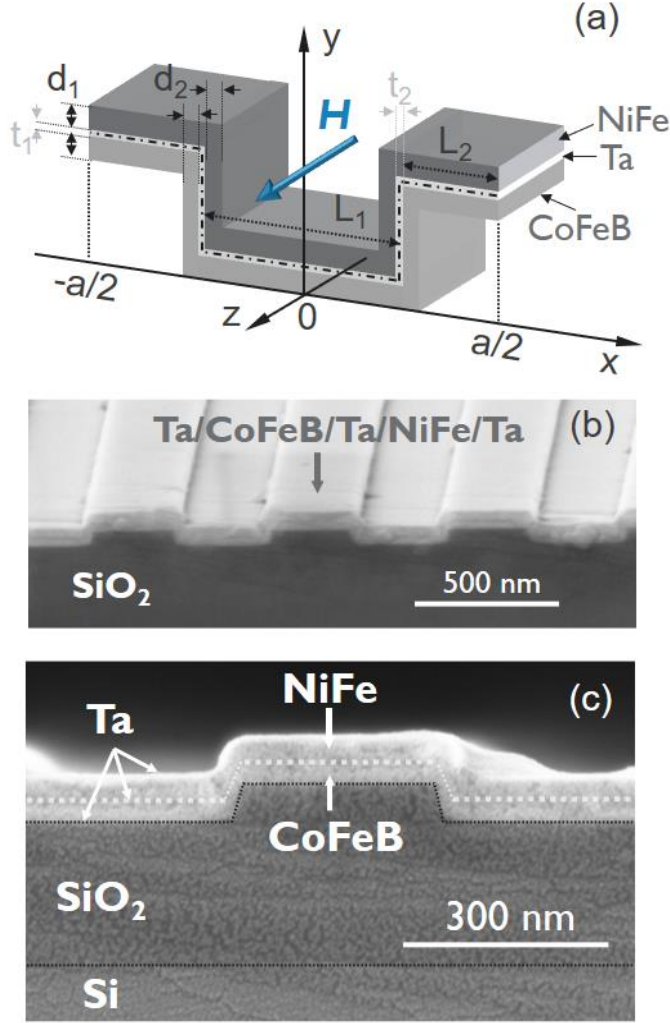

FIG. 1. (a) Schematic of the CoFeB/Ta/NiFe meander structure. The CoFeB and NiFe layers have identical thicknesses on both the horizontal ($d_1$) and vertical ($d_2$) segments. $t_1$ and $t_2$ denote the respective thicknesses of the Ta layer. (b) and (c) SEM images of the CoFeB/Ta/NiFe meander structures.

The SW dispersion was measured by BLS in the backscattering configuration using a Sandercock (3+3)-pass tandem Fabry-Perot interferometer.[18] An in-plane magnetic field of $m_0H = 50$ mT was applied along the groove length (z-direction) and perpendicular to the incidence plane of light (x-y plane), in the so-called magnetostatic surface wave (MSSW) configuration. BLS spectra were recorded by changing the incidence light angle from 0 to 60° in steps of 2°. This corresponds to a sweep of in-plane SW wavevector $k = (4\pi/\lambda)\times\sin(q)$ along the x-direction from 0 to $2.05\times10^7$ rad/m, with $\lambda = 532$ nm the wavelength of the laser.[19] The hysteresis loop (not shown), measured by vibrating sample magnetometry with the external magnetic field applied along the z-direction, was square with a coercive field of about 5 mT. No evidence of separate switching fields of the individual CoFeB and NiFe layers was found. The hysteresis loop further indicated that the sample magnetization was saturated at $m_0H = 50$ mT, the field used for the BLS measurements.

Figure 2(a) represents a sequence of BLS spectra for the CoFeB/Ta/NiFe sample measured at different incidence angles, *i.e.* at different wavevectors *k*. The spectra are characterized by a significant asymmetry of the peak intensity when comparing negative (Stokes) and positive (anti-Stokes) frequency shifts. Consequently, some peaks are only clearly visible at the Stokes side or *vice versa*. Three frequency regions could be identified, as marked by the blue (7.5 to 10.5 GHz), yellow (16.1 to 22.1 GHz), and green (25.3 to 28.0 GHz) areas in Fig. 2(a). The regions are characterized by peaks whose frequency position does not change over the range of explored wavevectors. A notable feature was the variation of the peak intensity as a function of *k*, which is a well-known effect for stationary modes and depends on the BLS cross-section dependence on the mode spatial profiles and their symmetry with respect to the incident plane of light.[20] For example, in the lowest frequency range (blue region), a triplet of modes was observed whose intensities change as a function of *k*. The peak at 7.52 GHz had the largest intensity at small wavevectors and decreased monotonically with *k*. By contrast, the peak at 8.78 GHz had maximum intensity around $k \approx 1.2\times10^7$ rad/m. In the intermediate frequency range between the blue and yellow regions, two peaks were detected that are marked by the red (blue) arrows in the Stokes (anti-Stokes) spectra. These peaks exhibited significant frequency variation with *k*, which indicates that the corresponding modes are dispersive. Furthermore, these two peaks showed opposite frequency dependences: the Stokes (anti-Stokes) peak frequency increased (decreased) up to $k = 1.18 \times10^7$ rad/m; however, the trend was subsequently reversed for at higher wavevectors.

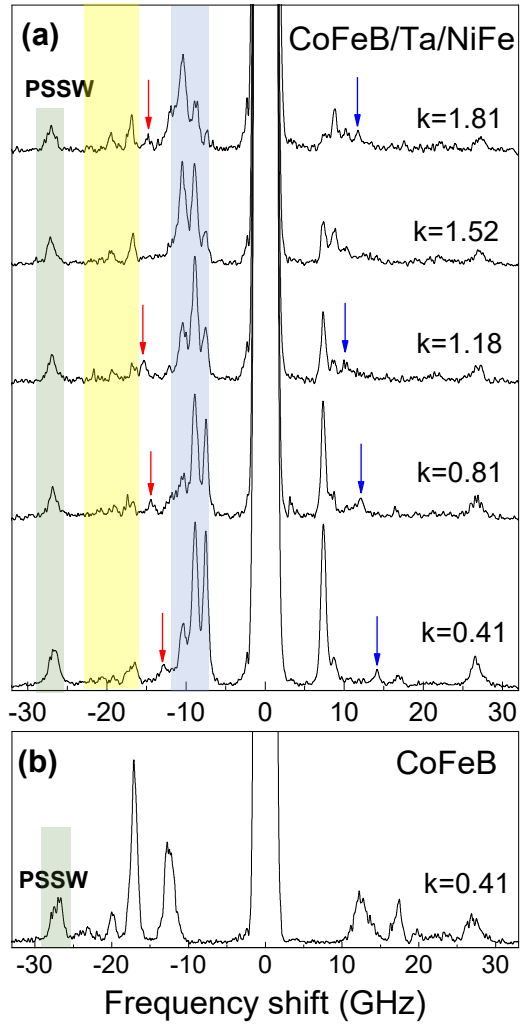

FIG. 2. Experimental BLS spectra for (a) CoFeB/Ta/NiFe and (b) CoFeB meander structures ($\mu_0 H_0 = 50$ mT) for different wavevectors (in $10^7$ rad/m). In (a), the blue, yellow, and green areas indicate frequency ranges that include peaks, whose frequencies does not change with wavevector. By contrast, red and blues arrows indicate peaks that exhibit sizeable frequency dispersion.

In Fig. 2 (b), the measured BLS spectrum at $k = 0.41 \times 10^7$ rad/m of a reference meander structure with a single CoFeB film is shown for comparison. Significant differences were observed between the spectra of the two samples in terms of mode frequencies and intensities. For example, there were no peaks below 12.5 GHz for the single-layer CoFeB sample, whereas the CoFeB/Ta/NiFe sample presented a triplet of peaks, as discussed above. This indicates that the triplet corresponds to either SW modes localized in the NiFe layer or to hybridized modes in the CoFeB/Ta/NiFe magnetic bilayer. Both samples display weakly dispersive peaks near 27 GHz, as seen in the green area of Figs.

2(a) and (b). The modes can thus be attributed to first-order perpendicular standing SW modes (PSSW) resonating across the thickness of the CoFeB film.

To model the magnonic band structure of the 3D MC, micromagnetic simulations were performed using the open-source GPU accelerated MuMax3 software.[21] The single-layer CoFeB and magnetic bilayer CoFeB/Ta/NiFe meander-shaped unit cell structure was discretized into cubic micromagnetic cells of dimensions $\Delta_x \times \Delta_y \times \Delta_z = 1 \times 1 \times 1$ nm$^3$. To calculate the SW dispersion relation, $N_p$ = 31 periods of the meander structure were simulated. Hence, the total length of the simulated structure in x-direction was $L = N_p \times a = 18.6$ µm. In the area $L \times s$, where $s = h+2d_1+t_1$, a rectangular grid with the size of $N_x \times N_y$ was specified, where $N_x = L/\Delta_x, N_y = L/\Delta_y$ are the number of grid nodes along the x and y axes, respectively. An out-of-plane sinc-shaped magnetic field with an amplitude of 1 mT and a cut-off frequency of $f_c$=27 GHz was applied in a 50 nm wide region in the center of the unit cell to excite spin waves. The sample was magnetized by a bias field of $m_0H = 50$ mT along the z-axis (see Fig. 1 (a)). A similar procedure was carried out in the time domain within a simulation time of T = 300 ns. The sampling time step was 9.26 ps. The spin-wave dispersion was then obtained by performing a 2D Fast Discrete Fourier Transform.[22] The magnetic materials were represented in the MuMax3 simulations by the following parameters: saturation magnetization $M_s$ (CoFeB) = 1275 kA/m, $M_s$ (Nife)= 795 kA/m, exchange constant $A_{ex}$ (CoFeB) = 1.43 × 10$^{-11}$ J/m, $A_{ex}$ (NiFe) = 1.11 × 10$^{-11}$ J/m.[21,23]

Figure 3 presents a comparison between measured and calculated dispersion relations for the CoFeB/Ta/NiFe and CoFeB meander structures, as well as the dispersion of an equivalent planar CoFeB/Ta/NiFe magnetic bilayer. In general, excellent agreement was observed between measured and calculated dispersion relations. For the CoFeB/Ta/NiFe meander structure (Fig. 3 (a)), the observed three modes below 11 GHz (labeled I, II, and III) and the two modes above 16 GHz (modes VI and VII) were dispersionless, i.e. their frequency was independent of the wavevector. By contrast, two modes (IV and V) exhibited an antiphase frequency oscillation with identical amplitude in the frequency range between 12 and 15.3 GHz. These two modes correspond to Block-type SWs that propagate over the entire sample. Mode crossing and the absence of a bandgap was observed at $k = \pi/a$. Furthermore, when compared to the behavior of the single-layer CoFeB MC,[17] the dispersive modes in the CoFeB/Ta/NiFe sample had smaller group velocity and a smaller magnonic bandwidth (3.3 GHz vs. 5.5 GHz).

The dispersion relation of the planar CoFeB/Ta/NiFe magnetic bilayer contains two modes, which due to dynamic dipolar coupling, correspond to acoustic (at a higher frequency) and optic modes (at a lower frequency) and are associated with the in-phase and out-of-phase precession of the dynamic magnetization in the two magnetic layers, respectively.[24] The acoustic mode is mainly localized in the bottom CoFeB layer whereas the optic branch possesses higher intensity in the NiFe layer. It is worth noting that the frequency of the dispersive modes with positive group velocity (*i.e.* modes IV) in the two meander structures depended on $k$ in similar manner than the optic modes in planar CoFeB/Ta/NiFe films (see Fig. 3(c)).

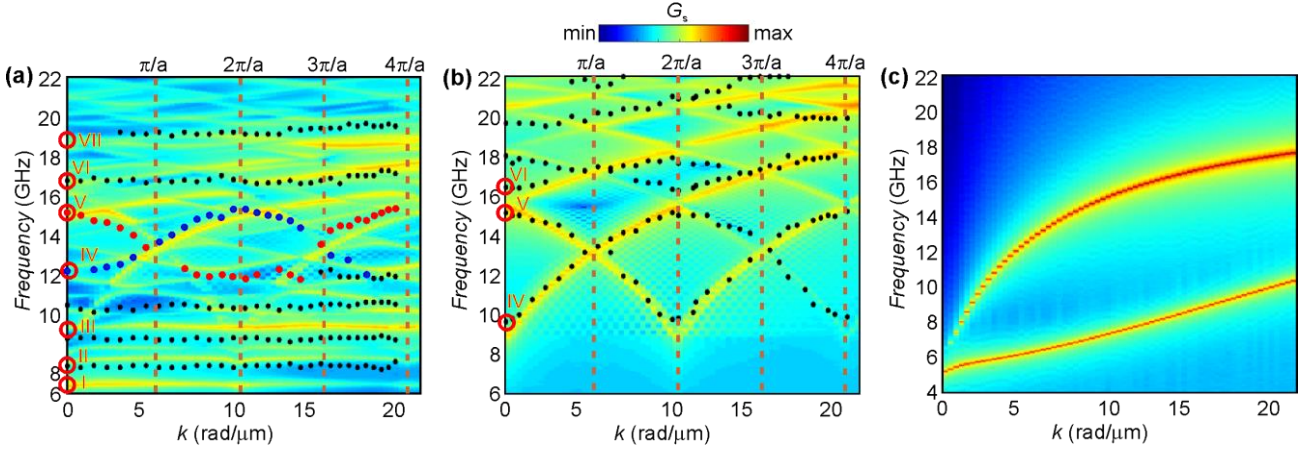

FIG. 3. Comparison between measured (points) and simulated (color map) dispersion relations for (a) CoFeB/Ta/NiFe and (b) CoFeB meander structures. (c) presents the calculated dispersion for flat (unpatterned) CoFeB/Ta/NiFe. In all cases, the external magnetic field was fixed at $\mu_0H$=50 mT, applied along the $z$-axis, while the in-plane SW wavevector $k$ was swept along the $x$-direction. The vertical dashed lines mark the edges of the BZs ($n\pi/a$, with $n$ = 1, 2, 3, and 4).

In the next step, the amplitude and the spatial localization of the modes in the two MCs were further investigated by analyzing the dynamic component of magnetization at $k = 0$ (*i.e.* in the center of the BZ), as shown in Fig. 4. For the CoFeB/Ta/NiFe meander structure (Fig. 4(a)), the calculated spatial distributions of both dispersive and nondispersive modes were symmetric with respect to the $y$-axis. Furthermore, the simulations show that the amplitude of modes I, II and III were mainly concentrated in the topmost NiFe layer with an increasing number of nodes in the horizontal segments and resulting in an increase of the mode frequency. We note that the frequencies of these modes are below the ferromagnetic resonance of a continuous CoFeB film (the saturation magnetization of NiFe is smaller than that of CoFeB), so that the magnetization precession in CoFeB is not resonantly excited but undergoes a forced oscillation.[25,26] The spin precession amplitude for this mode triplet in the vertical segments of the CoFeB and NiFe layers oscillates in antiphase (180° phase shift) due to the presence of interlayer dipolar coupling.

By contrast, the profile of mode IV was quasi-uniform within the CoFeB layer and showed negligible amplitude in NiFe. This mode had an extended spatial character and corresponded to the dispersive mode represented by the blue circles in Fig. 3(a). A similar distribution was observed also for the single-layer CoFeB MC (see Fig. 4(b)), even if the corresponding frequency of this mode was lower at around 10 GHz. Mode V of the CoFeB/Ta/NiFe MC also had an extended profile with the spin precession amplitude mainly concentrated in NiFe, although it was complemented by a sizeable in-phase amplitude in the CoFeB layer. This supports the propagating character of this mode, in agreement with the experimental results. Similar behavior was also observed for the single-layer CoFeB meander structure. Mode VI was mainly concentrated in the top NiFe and exhibited a large number of oscillations. This explains its relatively small peak intensity observed in the experimental spectra in Fig. 2(a).

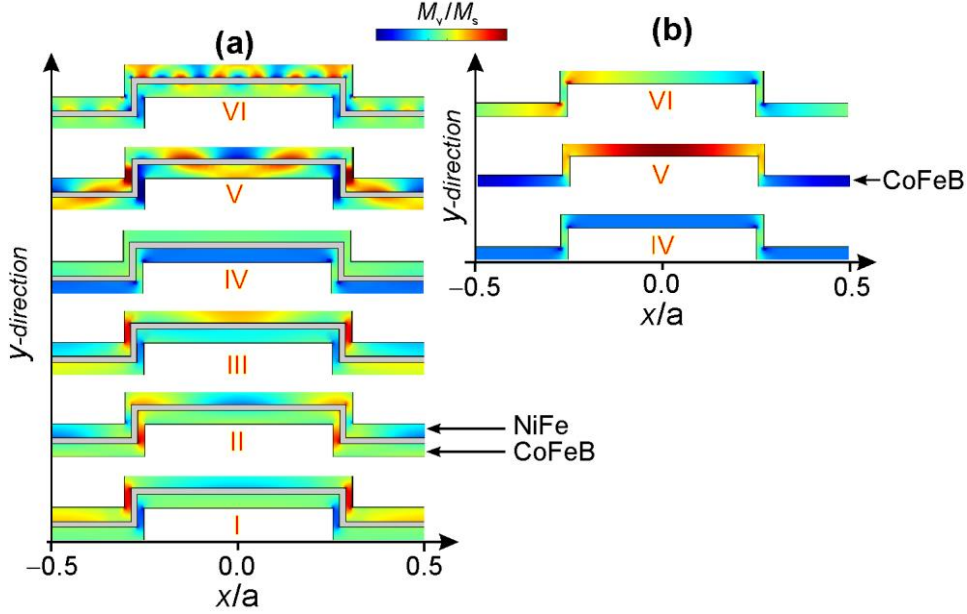

FIG. 4. Simulated spatial profiles (normalized out-of-plane component of dynamic magnetization, $M_y/M_s$) of the SW modes (a) I to VI for the CoFeB/Ta/NiFe meander structure and (b) IV, V, and VI for the single-layer CoFeB meander structure.

    In conclusion, we have studied both experimentally and numerically the SW dispersion relation of in-plane saturated CoFeB/Ta/NiFe meander-shaped MCs and compared the results to those of an equivalent sample comprising a single CoFeB layer. The magnonic band structures of the two samples are markedly different, both in terms of observed modes and in their dependences on the wavevector, *i.e.* whether they are stationary or dispersive in nature. A narrower width of the magnonic band has been observed for the CoFeB/Ta/NiFe structure than for the CoFeB sample. The properties of the individual modes have been further characterized by the phase relation (in-phase or out-of-phase) between the magnetization oscillations in the two layers and their localization in the horizontal

and vertical segments. The results show that meander structures containing layered magnetic structures can be considered as prototypes of 3D magnonic crystals that may be used as a basic element in complex multilevel magnonic waveguide networks.

Imec's contribution to this work has been supported by its industrial affiliate program on beyond-CMOS logic as well as by the European Union's Horizon 2020 research and innovation program within the FET-OPEN project CHIRON under grant agreement No. 801055. The numerical simulation and theoretical model for three-dimensional magnonic band structure has been supported by the Russian Science Foundation (Project No. 20-79-10191). E.B. acknowledges support from the Russian Ministry of Education and Science (Project No. FSRR-2020-0005) and RFBR (Project No.19-29-03034). S. N. acknowledges support by the Russian Science Foundation (Project 19-19-00607). The authors would like to thank Danny Wan, Anshul Gupta, Shreya Kundu, and Robert Carpenter at imec for support of the sample fabrication.

REFERENCES

[1] J. Topp *et al.*, Phys. Rev. Lett. **104**, 207205 (2010).

[2] G. Gubbiotti, X. Zhou, Z. Haghshenasfard, M. G. Cottam and A.O. Adeyeye, Phys. Rev. B **97**, 134428 (2018).

[3] D. Kumar, J. W. Klos, M. Krawczyk and A. Barman *J. Appl. Phys.* **115**, 043917 (2014).

[4] A. Barman, S. Mondal, S. Sahoo, and A. De, J. Appl. Phys. **128**, 170901 (2020).

[5] M. Krawczyk and D. Grundler, J. Phys.: Condens. Matter **26**, 123202 (2014).

[6] S. Tacchi, G. Gubbiotti, M. Madami, and G. Carlotti, J. Phys.: Condens. Matt. **29**, 073001 (2017).

[7] B. Obry, P. Pirro, T. Brächer, A.V. Chumak, J. Osten, F. Ciubotaru, A.A. Serga, J. Fassbender, and B. Hillebrands, Appl. Phys. Lett. **102**, 202403 (2013).

[8] M. Inoue, A. Baryshev, H. Takagi, P. B. Lim, K. Hatafuku, J. Noda, and K. Togo, Appl. Phys. Lett. **98**, 132511 (2011)

[9] H. Merbouche, M. Collet, M. Evelt, V. E. Demidov, J. L. Prieto, M. Munoz, J. B. Youssef, G. de Loubens, O. Klein, S. Xavier, O. D'Allivy Kelly, P. Bortolotti, V. Cros, A. Anane, and S. O. Demokritov, ACS Applied Nano Materials **4**, 1 121 (2021).

[10] A. V. Chumak, A. A. Serga, and B. Hillebrands, Nat. Commun. **5**, 4700 (2014).

[11] L. Brunet et al., IEEE Symposium on VLSI Technology, Honolulu, HI, USA, 2016, 1 (2016).

[12] A. Vandooren et al., IEEE Symposium on VLSI Technology, Honolulu, HI, 2018, 69 (2018).

[13] G. Gubbiotti, Three-Dimensional Magnonics (Jenny Stanford, Singapore, 2019).

[14] V. K. Sakharov, E. N. Beginin, Y. V. Khivintsev, A. V. Sadovnikov, A. I. Stognij, Y. A. Filimonov, and S. A. Nikitov, Appl. Phys. Lett. **117**, 022403 (2020).

[15] A. A. Martyshkin, E. N. Beginin, A. I. Stognij, S. A. Nikitov, and A. V. Sadovnikov, IEEE Magn. Lett. **10**, 5511105 (2019).

[16] E. N. Beginin, A. V. Sadovnikov, A. Yu. Sharaevskaya, A. I. Stognij, and S. A. Nikitov, Appl. Phys. Lett. **112**, 122404 (2018).

[17] G. Gubbiotti, A. Sadovnikov, E. Beginin, S. Nikitov, D. Wan, A. Gupta, S. Kundu, G. Talmelli, R. Carpenter, I. Asselberghs, I. P Radu, C. Adelmann, and F. Ciubotaru, Phys. Rev. Appl. **15**, 014061 (2021).

[18] J. R. Sandercock, in *Light Scattering in Solids III*, edited by M. Cardona and G. Guntherodt, Springer Series in Topics in Applied Physics Vol. 51, Springer-Verlag, Berlin, (1982), p. 173.

[19] G. Carlotti and G. Gubbiotti, J. Phys.: Condens. Matter **14**, 8199 (2002).

[20] J. Jorzick, S. O. Demokritov, C. Mathieu, B. Hillebrands, B. Bartenlian, C. Chappert, F. Rousseaux and A. N. Slavin, Phys. Rev. B 60, 15194 (1999).

[21] A. Vansteenkiste, J. Leliaert, M. Dvornik, M. Helsen, F. Garcia-Sanchez, and B. Van Waeyenberge, AIP Advances **4**, 107133 (2014).

[22] A. V. Sadovnikov, S. A. Odintsov, E. N. Beginin, A. A. Grachev, V. A. Gubanov, S. E. Sheshukova, Yu. P. Sharaevskii, S. A. Nikitov, JETP Letters **107**, 25–29 (2018).

[23] A. Vansteenkiste A., B. Van De Wiele, J. Magn. Magn. Mater. **323**, 2585 (2011).

[24] G. Gubbiotti, M. Kostylev, N. Sergeeva, M. Conti, G. Carlotti, T. Ono, A. N. Slavin and A. Stashkevich, Phys. Rev. B **70**, 224422 (2004).

[25] M. L. Sokolovsky and M. Krawczyk, J. Nanopart. Res. **13** 6085 (2011).

[26] G. Gubbiotti, S. Tacchi, M. Madami, G. Carlotti, S. Jain, A. O. Adeyeye, and M. P. Kostylev, Appl. Phys. Lett. **100**, 162407 (2012).